\documentclass[prb,twocolumn,amsmath,amssymb,showpacs]{revtex4-1} 
\pdfoutput=1
\usepackage{graphics}
\usepackage{dcolumn}
\usepackage{bm}
\usepackage{color}
\usepackage{epstopdf}
\usepackage{epsfig}

\newcommand{\spin}{\sigma}
\newcommand{\ospin}{\overline{\sigma}}
\newcommand{\kv}{{\bf k}}
\newcommand{\qv}{{\bf q}}
\newcommand{\Qv}{{\bf Q}}
\newcommand{\Nav}{\langle n \rangle }

\begin{document}

\title{Pairing symmetry of the one-band Hubbard model in the paramagnetic weak-coupling limit: a numerical RPA study}

\author{A. T. R\o mer$^{1}$, A. Kreisel$^1$,  I. Eremin$^2$, M. A. Malakhov$^{3}$, T. A. Maier$^4$, P. J. Hirschfeld$^{5}$, B. M. Andersen$^{1}$}
\affiliation{$^1$Niels Bohr Institute, University of Copenhagen, DK-2100 Copenhagen, Denmark\\
$^2$Institut f\"ur Theoretische Physik III, Ruhr-Universit\"at Bochum, D-44801 Bochum, Germany\\
$^3$ Institute of Physics, Kazan (Volga Region) Federal University, 420008 Kazan, Russian Federation\\
$^4$ Center for Nanophase Materials Sciences, Oak Ridge National Laboratory, Oak Ridge, Tennessee 37831, USA\\
$^5$ Department of Physics, University of Florida, Gainesville, Florida 32611, USA
}
\begin{abstract}
We study the spin-fluctuation-mediated superconducting pairing gap in a weak-coupling approach to the Hubbard model for a two dimensional square lattice in the paramagnetic state.
Performing a comprehensive theoretical study of the phase diagram as a function of filling, we find that the superconducting gap exhibits transitions from $p$-wave at very low electron fillings to $d_{x^2-y^2}$-wave symmetry close to half filling in agreement with previous reports. At intermediate filling levels, different gap symmetries appear as a consequence of the changes in the Fermi surface topology and the associated structure of the spin susceptibility. In particular, the vicinity of a van Hove singularity in the electronic structure close to the Fermi level has important consequences for the gap structure in favoring the otherwise sub-dominant triplet solution over the singlet $d$-wave solution. By solving the full gap equation, we find that the energetically favorable triplet solutions are chiral and break time reversal symmetry. Finally, we also calculate the detailed angular gap structure of the quasi-particle spectrum, and show how spin-fluctuation-mediated pairing leads to significant deviations from the first harmonics both in the singlet $d_{x^2-y^2}$ gap as well as the chiral triplet gap solution.
\end{abstract}


\maketitle

\section{INTRODUCTION}

The Hubbard model for electrons in metals is considered by many to contain the essential
ingredients for high-temperature unconventional superconductivity.~\cite{PWA,Scalapino_review,Scalapino86,Emery,Miyake} Most recently, it was also realized with ultracold atoms
in optical lattices.~\cite{Koel_05,Jordens_08,Schneider_08} In contrast to bulk materials, the interaction strength and e.g. the concentration of fermions can be changed relatively easily in optical lattice systems, allowing for systematic studies of the regimes of doping also far away from half filling (one electron per site).  Theoretically, there has been significant progress in understanding unconventional superconducting instabilities driven by repulsive interactions, particularly in the weak-coupling limit,~\cite{kagan,zanchi,metzner,raghu_10} 
and within various numerical techniques discussed in Ref.~\onlinecite{Scalapino07} and more recently in Refs. ~\onlinecite{gull,Deng14,Staar14,Chan15}. 
At present, however, the superconducting phase diagram of the two dimensional Hubbard model remains largely unknown, especially at intermediate fillings and at low temperatures.

Superconductivity mediated by antiferromagnetic (AF) spin fluctuations near half filling was studied e.g. in the early work of Scalapino {\it et al}.\cite{Scalapino86} for a three dimensional paramagnetic system close to an AF instability. There it was found that the Coulomb repulsion between two electrons may give rise to a substantial Cooper-pairing strength due to the proximity of the AF instability. In this case, the superconducting gap symmetries are intimately related to the structure of the paramagnetic spin susceptibility, and therefore sensitive to the geometry of the Fermi surface. In the work of Scalapino {\it et al.},\cite{Scalapino86} it was found that the $d_{x^2-y^2}$ solution is favored close to half filling as a consequence of the spin susceptibility peak at the AF wave vector $\Qv=(\pi,\pi,\pi)$, whereas the limit of very small electron filling prefers a spin triplet $p$-wave solution due to a susceptibility peak near $\qv=(0,0,0)$. These arguments also carry over to two dimensions, where the corresponding wave vectors are $(\pi,\pi)$ and $(0,0)$, respectively.
Since the seminal work of Ref.~\onlinecite{Scalapino86}, spin-fluctuation pairing in the two dimensional Hubbard model has been studied extensively by analytical and numerical approaches, and the sensitivity to the underlying Fermi surface gives rise to a particularly rich phase diagram with many emerging gap symmetries. In particular, it has been noted that there is support for triplet superconductivity in a quite large region of doping in the limit of small $U$.~\cite{Hlubina99} In Ref.~\onlinecite{raghu_10} it was found that the vicinity of a van Hove singularity near the Fermi level causes enhancement of the triplet superconductivity compared to singlet solutions for sizable values of nearest-neighbor hopping integrals. Furthermore, it was found that the superconducting phase diagram in the single-band Hubbard model (with $t'=0$) is quite robust against an inclusion of the long-range Coulomb interaction at least in the weak-coupling limit. \cite{Raghu2012}

In fact, the dominant interest of the community regarding possible superconducting channels and angular variations of the superconducting gaps on the Fermi surface was often restricted to the doping near the half-filling because of the high-T$_c$ cuprates. In this regard, there is a consensus that the dominating gap symmetry of moderately hole- or electron-doped cuprates is of the $d_{x^2-y^2}$-wave form.\cite{Armitage10,Hashimoto14} Many experiments, however, find evidence that the quasi-particle gap does not always follow the simplest lowest order harmonic form as given by $\Delta_\kv=\Delta [\cos(k_x)-\cos(k_y)]$; the gap may exhibit its maximum value not at the antinodes, as would be the case for the leading harmonic $d$-wave form, but at a different location on the Fermi surface.
In the electron-doped compounds, such a non-monotonic $d$-wave form was observed e.g. in Raman spectroscopy on NCCO~\cite{Blumberg02} and in ARPES experiments on PLCCO.~\cite{Matsui05} In these cases, the position of the maximum gap value was related to the position of the so-called hot spots, i.e. segments of the Fermi surface which are connected by $\Qv=(\pi,\pi)$. This points towards a pairing interaction mediated by AF spin fluctuations, in which case we expect the dominating pairing strength at $\Qv$, and consequently the largest gaps to be located at pairs of $\kv$ and $\kv'$ on the Fermi surface separated by $\Qv$. In the case of hole doped cuprates, ARPES experiments report enhancement of the gap in the antinodal regions.\cite{Hashimoto14}  This has been attributed to the presence of the pseudo-gap since mainly cuprates in the underdoped regime display a significant non-monoticity of the observed gap, and the antinodal gap is known to persist above $T_c$.~\cite{Hashimoto14} However, one report on La$_{2-x}$Sr$_{x}$CuO$_{4}$ with hole doping of 15 \%,~\cite{Terashima07} close to optimal doping, found a strong deviation from the [$\cos(k_x)-\cos(k_y)$] form. This indicates that the non-monotonicity may not be entirely caused by the pseudo-gap, and it is possible that spin-fluctuation pairing effects also cause gap enhancements near the antinodal regions of hole-doped cuprates.

Quite generally, the interplay of spin fluctuations, superconducting gap and the underlying Fermi surface topology is an interesting characteristic of spin fluctuations mediated Cooper-pairing within the weak-coupling approach. In this regard, it is essential to study their evolution not only near half-filling but also for the entire phase diagram of the single-band Hubbard model. Despite previous efforts in this direction\cite{Hlubina99,raghu_10,Deng14,Chubukov92,Katanin2003}, the systematic knowledge on the weak-coupling superconducting phase diagram for the single-band Hubbard model is still missing, especially for the intermediate doping range and $t'/t>0.5$.

In this paper, we investigate the spin-fluctuation-mediated pairing interaction within the RPA for a large set of possible singlet and triplet solutions to the gap equations within a weak-coupling approach for the entire doping range and strong asymmetry between electron and hole doping. Concentrating on the less explored intermediate doping range, we show how the shape and topology of the Fermi surface play a decisive role for the final preferred gap symmetry. In particular, we find that the change of the Fermi surface topology, associated with the chemical potential crossing the van Hove singularity with a logarithmic divergence in the density of states, has a strong effect on the potential gap solutions in various symmetry channels, and favors a higher order triplet gap over the singlet $d_{x^2-y^2}$ solution even at significant electron filling. We study how the different gap symmetry solutions evolve as a function of filling and next-nearest neighbor hopping integral $t'/t$, and map out the detailed gap structure arising directly from the spin-fluctuation pairing mechanism. This includes deviations from the [$\cos(k_x)-\cos(k_y)$] form of the superconducting gap in the singlet channel close to half filling, as well as the form of the higher order triplet gap. In the triplet channel we find that the preferred solutions are those that break time reversal symmetry.

\section{MODEL AND METHOD}
We consider the Hubbard model for a two-dimensional square lattice
\begin{equation}
 H=\sum_{\kv \spin}\xi_kc_{\kv\spin}^\dagger c_{\kv\spin}+\frac{U}{2N}\sum_{\kv,\kv',\qv}\sum_{\spin}c_{\kv'\spin}^\dagger c_{-\kv'+\qv\ospin}^\dagger c_{-\kv+\qv\ospin} c_{\kv\spin},
\end{equation}
where $\xi_\kv=-2t[\cos(k_x)+\cos(k_y)]-4t'\cos(k_x)\cos(k_y)-\mu$ with $t$ being the hopping integral to nearest neighbors, and $t'<0$ the hopping integral between next-nearest neighbors. In the following we set $t=1$ and restrict ourselves to the case of negative values of $t'$.

A spin-fluctuation-mediated interaction can combine two electrons of opposite spin or the same spin into a Cooper pair. The pairing interaction is derived from higher order diagrams of the repulsive Coulomb interaction $U$.~\cite{berkschrie,Scalapino86} In the case of opposite electron spins, the diagrams consist of an even number of bubbles as well as ladder diagrams, which correspond to spin preserving or spin flip interactions, respectively. For same spin electrons, the interaction is derived from an odd number of bubble diagrams, and in this case only spin preserving interactions are allowed. Specifically, the interactions are given by~\cite{Scalapino86}
\begin{eqnarray}
 \Gamma_{\kv,\kv'}^\textrm{opp.sp}&=&U+\frac{U^2}{2}\chi_{(\kv-\kv')}^{sp}-\frac{U^2}{2}\chi_{(\kv-\kv')}^{ch}+U^2\chi_{(\kv+\kv')}^{sp},  \label{eq:PMoppspin} \nonumber\\
&& \\
 \Gamma_{\kv,\kv'}^\textrm{same sp}&=&-\frac{U^2}{2}\chi_{(\kv-\kv')}^{sp}-\frac{U^2}{2}\chi_{(\kv-\kv')}^{ch},  \label{eq:PMsamespin}
 \end{eqnarray}
 with the spin and charge susceptibilities given by
 \begin{eqnarray}
 \chi^{sp}_\qv&=&\frac{\chi_0(\qv)}{1-\bar{U}\chi_0(\qv)}, \label{eq:chisp}\\
 \chi^{ch}_\qv&=&\frac{\chi_0(\qv)}{1+\bar{U}\chi_0(\qv)}.\label{eq:chich}
\end{eqnarray}
Here $\bar{U}=U/z$ is a renormalized Coulomb interaction. In the main part of the paper we use $z=1$ corresponding to the usual RPA, but when solving the full gap equation we use a renormalization ($z=2,3$) to achieve a larger pairing strength for numerical convergence.
Equations~(\ref{eq:PMoppspin}) and~(\ref{eq:PMsamespin}) provide a measure of the interaction strength, and we neglect the energy dependence of the interactions.
The bare susceptibility in the paramagnetic phase is given by the Lindhard function
\begin{equation}
 \chi_0(\qv,\omega)=\frac{1}{N}\sum_\kv\frac{f(\xi_{\kv+\qv})-f(\xi_{\kv})}{\omega+\xi_\kv-\xi_{\kv+\qv}+i\eta},
\end{equation}
which is evaluated a zero energy ($\omega=0$).
The gap equation arises from a standard mean-field decoupling of the interaction Hamiltonian. In the singlet $(s)$ and triplet $(t)$ channel it takes the form
\begin{eqnarray}
  \Delta^{s/t}_\kv&=&
  -\frac{1}{2N}\sum_{\kv'}\Gamma^{s/t}_{\kv,\kv'}\frac{\Delta^{s/t}_{\kv'}}{2E^{s/t}_{\kv'}}\tanh\Big(\frac{E^{s/t}_{\kv'}}{2k_BT}\Big),\nonumber\\
&&\label{eq:SCGapEquationNS}
\end{eqnarray}
with
\begin{eqnarray}
E_\kv&=&\sqrt{\xi_{\kv}^2+|\Delta^{s/t}_\kv|^2}.
\end{eqnarray}
In the calculation of the superconducting gap, the potential forms stated in Eqs.~(\ref{eq:PMoppspin}) and ~(\ref{eq:PMsamespin}) must be symmetrized or antisymmetrized with respect to momentum in the singlet and triplet channel, respectively.
In case of singlet pairing, the interaction potential $\Gamma_{\kv,\kv'}$ is given by the opposite spin vertex, as stated in Eq.~(\ref{eq:PMoppspin}), and we have
\begin{equation}
\Gamma^{s}_{\kv,\kv'}=\Gamma^{\rm opp. sp}_{\kv,\kv'}+\Gamma^{\rm opp. sp}_{-\kv,\kv'}.
\end{equation}
For triplet pairing, the same effective interaction is obtained irrespective of whether the same spin or opposite spin interaction vertex is used
\begin{equation}
\Gamma^{t}_{\kv,\kv'}=\Gamma^{\rm opp. sp}_{\kv,\kv'}-\Gamma^{\rm opp. sp}_{-\kv,\kv'}=\Gamma^{\rm same~sp}_{\kv,\kv'}-\Gamma^{\rm same~sp}_{-\kv,\kv'}
\end{equation}
This is as expected in the paramagnetic phase. Note that the potential entering Eq.~(\ref{eq:SCGapEquationNS}) appears in the singlet (even in $\kv$) and triplet (odd in $\kv$) form explicitly. This symmetry directly carries over to the gap, ensuring that $\Delta_{\kv}^s=\Delta_{-\kv}^s$ and $\Delta_\kv^t=-\Delta_{-\kv}^t$.
When solving the gap equation as stated in Eq.~(\ref{eq:SCGapEquationNS}), we invoke a small energy cut-off around the Fermi surface, $\epsilon_c$, and allow for Cooper pair formation of all electronic states with $\xi_\kv \in [-\epsilon_c,\epsilon_c]$.

\begin{figure}[t!]
 \centering
  	\includegraphics[angle=0,width=0.99\linewidth]{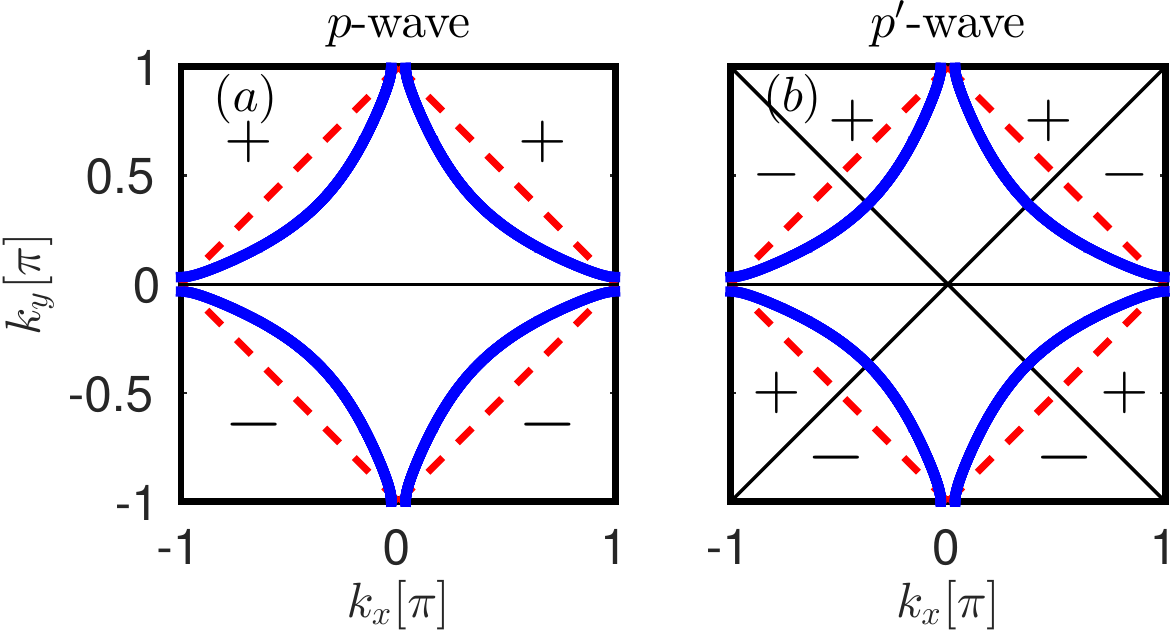}
  	\caption{(Color online) Illustration of the triplet gaps $p$-wave (a) and $p'$-wave (b). The latter is favored in the doping regime where the spin susceptibility has a peak or plateau at $\Qv=(\pi,\pi)$. The magnetic zone boundary is indicated by the dashed red line. Nodal lines in the two cases are shown by full black lines.}
  	\label{fig:tripletgaps}
 \end{figure}

In order to capture the filling dependence of the most prominent gap candidates, we project the pairing potential onto the leading order gap harmonics given by
\begin{subequations}
\begin{align}
 s^*&=\cos(k_x)+\cos(k_y) \label{eq:s}, \\
 d_{x^2-y^2}&=\cos(k_x)-\cos(k_y) \label{eq:d2}, \\
 d_{xy}&=\sin(k_x)\sin(k_y) \label{eq:dxy},\\
 g&=[\cos(k_x)-\cos(k_y) ]\sin(k_x)\sin(k_y), \label{eq:gxy}\\
 p&=\sin(k_x)\label{eq:p},\\
 p'&=[\cos(k_x)-\cos(k_y)]\sin(k_x).\label{eq:f}
\end{align}
\end{subequations}
Note that in the literature on Sr$_2$RuO$_4$ the $p'$ solution is sometimes called $f_{x^2-y^2}$-wave.~\cite{EreminEPL02}
The two triplet solutions $p$ and $p'$ belong to the same two-dimensional $E_u$ group, and are visualized in Fig.~\ref{fig:tripletgaps}.
We follow the procedure of Scalapino {\it et al.}~\cite{Scalapino86} and calculate the projection of the interaction vertex onto the basis functions

\begin{eqnarray}
\bar{\lambda}_\alpha&=&-\int_{FS}\frac{d\kv}{|v_\kv|} \int_{FS}\frac{d\kv'}{|v_\kv'|}g_\alpha(\kv)\Gamma^{s/t}_{\kv,\kv'}g_\alpha(\kv')/\int_{FS}\frac{d\kv}{|v_\kv|g^2_\alpha(\kv) },\nonumber \\
\label{eq:lam}
\end{eqnarray}
with $g_\alpha$ being one of the functions stated in Eqs.~(\ref{eq:s})-(\ref{eq:f}).
This procedure does not include possible higher order solutions. To determine these, we also solve the linearized gap equation in the singlet and triplet channel
\begin{equation}
\Big[ -\frac{1}{2(2\pi)^2}\int_{FS}\frac{d\kv'}{|v_{\kv'}|}\Gamma^{s/t}_{\kv,\kv'}\Big]\Delta_{\kv'}= \lambda_\kv \Delta(\kv),
\label{eq:lge}
\end{equation}
by diagonalization of the matrix
\begin{equation}
M_{\kv,\kv'}= -\frac{1}{2(2\pi)^2}\frac{l_{\kv'}}{|v_{\kv'}|}\Gamma^{s/t}_{\kv,\kv'}.
\end{equation}
Here $\kv$ and $\kv'$ are located on the Fermi surface and $l_\kv$ is the length of the Fermi surface segment associated with the point $\kv$. By this procedure we identify the leading instability as a function of electron filling and next-nearest neighbor hopping constant, $t'$.
We characterize the leading singlet solution according to its transformation properties into one of the four singlet representations $A_{1g}: s^*$, $B_{1g}:d_{x^2-y^2}$, $A_{2g}: g$, $B_{2g}:d_{xy}$. Note that the square lattice has one class of triplet solution ($E_u$) of which the $p$-wave is the lowest harmonic. However, in general we find the leading triplet solution to be higher order, as discussed in detail below.

\begin{figure*}[t!]
 	\includegraphics[angle=0,width=0.99\textwidth]{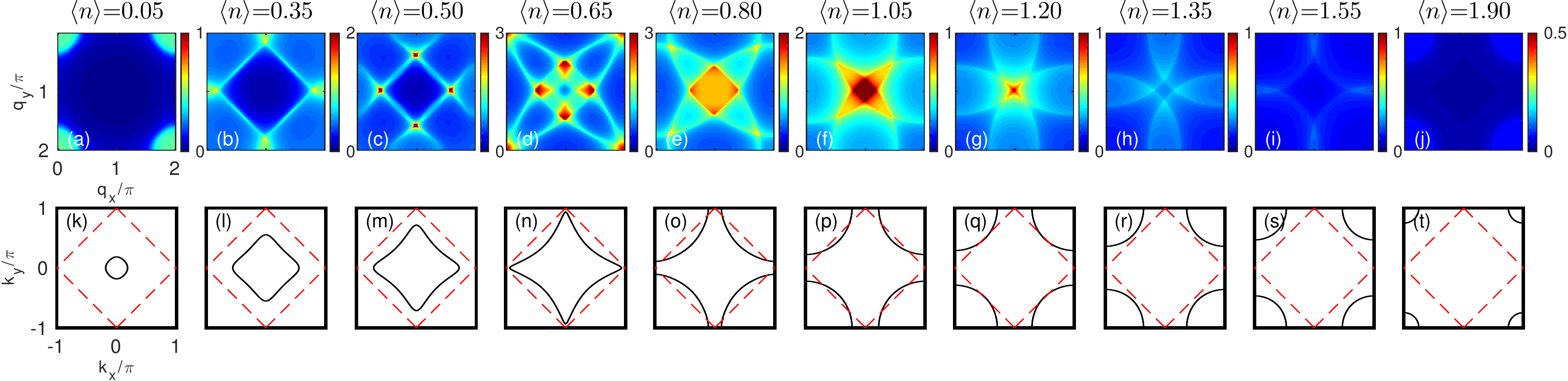}
 	\includegraphics[angle=0,width=0.95\textwidth]{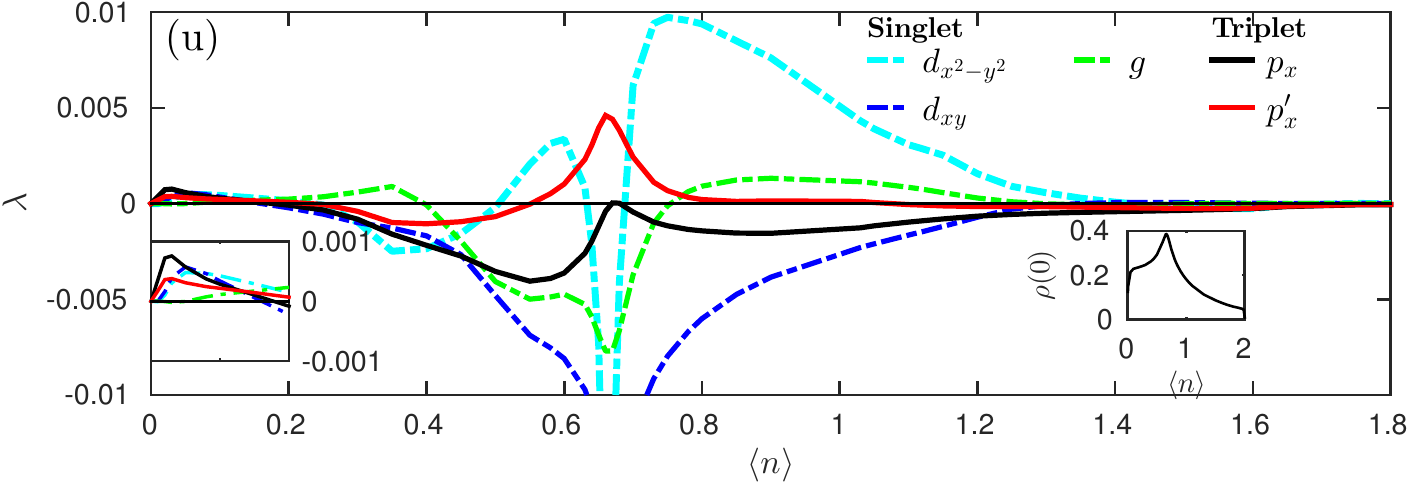}
	\caption{(Color online) (a-j) Spin susceptibility in the RPA approximation and (k-t) Fermi surfaces for fillings $\Nav = 0.05, 0.35, 0.50, 0.65, 0.80$ and $\Nav = 1.05, 1.2,1.35, 0.55, 0.90$  with $t'=-0.35$ and $U=1.75$. The AF zone boundary is shown by the dashed red line. Note the different colorbar range for each susceptibility plot.
	(u) Phase diagram for $\lambda$ as stated in Eq.~(\ref{eq:lam}) for a band with $t'=-0.35$ and $U=1.75$. The temperature is $k_BT=0.015$. Projection onto $s*$-wave ($\cos k_x +\cos k_y$) gives a negative value of $\lambda$ at all fillings. The left inset shows a zoom for low filling values $\Nav=0-0.2$.
	The inset to the right shows the density of states at the Fermi level, $\rho(0)$, as a function of filling.
	Note that at the filling for which $\rho(0)$ is maximal the triplet $p'$ solution is the leading instability. This we refer to as the van Hove critical density. For $t'=-0.35$ the van Hove critical density is $\langle n \rangle_{vH}$=0.66.}
 	\label{fig:lgeProjected0p35}
\end{figure*}

\begin{figure}[t!]
\centering
 	\includegraphics[angle=0,width=0.5\textwidth]{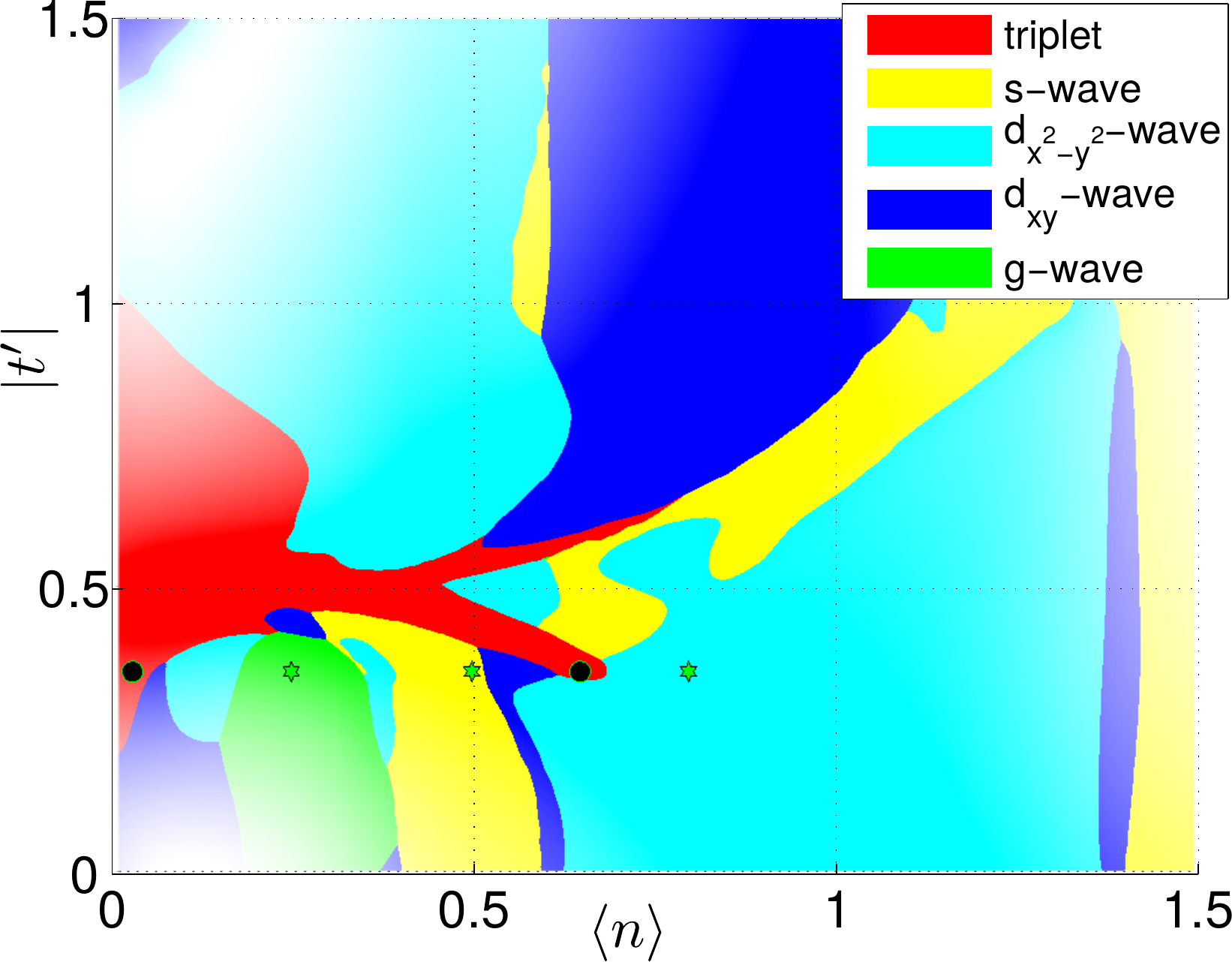}
 	\caption{(Color online) Phase diagram of the leading superconducting instability as a function of electron filling $\Nav$ and next-nearest neighbor hopping constant $|t'|$. Throughout, the value of $U$ is adjusted such that the leading eigenvalue is $\lambda=0.1$. For values of $U<3$, the eigenvalue is shown in full color. For large $U$ values in the range $3-8$ the eigenvalue is shown with brighter color in order to indicate this. We include only fillings $\Nav \leq 1.5$ since the instability is negligible for higher electron dopings for $U<8$.
 	The three green stars mark the positions for which we plot the singlet solutions in Figs.~\ref{fig:lged2} and~\ref{fig:lgesing}, and the two black filled circles mark the position for which we solve the full (non-linearized) gap equation in the triplet channel, Fig.~\ref{fig:TripletGapVsPhi}.}
 	\label{fig:tp-xdiagram}
\end{figure}

The linearized gap equation does not allow for a determination of complex gap solutions, which are time reversal symmetry broken (TRSB) solutions.
TRSB solutions can lead to a removal of gap nodes from the Fermi surface whereby there is a gain in condensation energy.
In the triplet channel, where solutions are doubly degenerate, TRSB solutions might be favored due to this effect. Therefore, we also address the non-linearized gap equation as stated in Eq.~(\ref{eq:SCGapEquationNS}) and show that the full self-consistent calculation finds that the triplet solutions are TRSB. Finally the solution of the full gap equation is used to obtain the angular structure of the resulting gap which may exhibit significant changes from the standard lowest harmonic due to the details of the momentum structure of the spin susceptibility.

\section{RESULTS}
\label{section_res}

\subsection{Doping dependence of the lowest harmonic solutions}
First we consider the Fermi surfaces and spin susceptibilities throughout the entire doping range for a next-nearest hopping integral $t'=-0.35$ relevant for cuprates.
As opposed to the early work in Ref.~\onlinecite{Scalapino86}, which reported the $t'=0$ case, we pay special attention to the electron-hole asymmetric case when the next-nearest neighbor hopping $t'$ is non-zero.
In Fig.~\ref{fig:lgeProjected0p35}(k-t) the Fermi surfaces at different electron fillings are shown. Note the transition of the Fermi surface between Fig.~\ref{fig:lgeProjected0p35}(n) and Fig.~\ref{fig:lgeProjected0p35}(o). In the latter case, the Fermi surface has "split up" at the antinodal positions $(\pi,0)$ and $(0,\pi)$. This splitting occurs when the van Hove singularities at $(\pm \pi,0)$ and $(0,\pm \pi)$ cross the Fermi level for a chemical potential of $\mu=4t'$. For $t'=-0.35$ this happens at the van Hove critical density, $\Nav_{vH}=0.66$.

The spin susceptibilities at the corresponding doping levels are shown in  Fig.~\ref{fig:lgeProjected0p35}(a-j). At very large dopings the susceptibility exhibits a broad peak around $\qv=(0,0)$ as seen in Figs.~\ref{fig:lgeProjected0p35}(a) and \ref{fig:lgeProjected0p35}(j) which develops into peaks at $(\pm\pi,0)$ and $(0,\pm\pi)$ as the doping is decreased as seen for $\Nav=0.35$ in Fig.~\ref{fig:lgeProjected0p35}(b).
 At intermediate doping levels the peaks at $(\pm\pi,0)$, $(0,\pm\pi)$ move inwards, and develop into the well-known quartet of incommensurate peaks at $(\pi\pm\delta,\pi)$ and $(\pi,\pi\pm\delta)$ as shown in Fig.~\ref{fig:lgeProjected0p35}(c), with $\delta$ decreasing as the system gets closer to half filling. A special feature is observed for fillings close to the van Hove critical density as seen from Fig.~\ref{fig:lgeProjected0p35}(d) where a $\qv=(0,0)$ peak develops as a direct consequence of the large density of states at the Fermi level. Close to half filling, a clear peak around $\Qv=(\pi,\pi)$ develops as shown Fig.~\ref{fig:lgeProjected0p35}(e-g).
Finally, at very large fillings the spin susceptibility becomes almost featureless as seen from Fig.~\ref{fig:lgeProjected0p35}(h-j).

In order to map out the filling dependence of the gap symmetries as defined in Eqs.~(\ref{eq:s})-(\ref{eq:f}) we project the pairing potential onto these symmetries as stated in Eq.~(\ref{eq:lam}). In the following, we do not show the $s^*$-wave results since these are highly suppressed at all doping levels.
We have chosen the Coulomb interaction $U=1.75$. While this choice leads to small values of $\lambda$, it allows us to avoid the instability to long range magnetic order over the entire phase diagram.
At very large hole and electron dopings, the peaks in the spin susceptibility are weak in intensity and, as a direct consequence, the projected pairing strengths are relatively small compared to half filling. This doping regime has a leading triplet $p$-wave solution, which for the hole doped case is visible from the left inset of Fig.~\ref{fig:lgeProjected0p35}. This is in agreement with the result of the $t'=0$ band.~\cite{Scalapino86}
In the filling regime $\Nav \sim 0.2 - 0.4$, the $g$-wave solution is favored, whereas none of the lowest harmonic solutions are supported in the filling regime around $\Nav=0.45$. It is important to note that the projection method does not take into account possible higher order solutions. These might dominate in some regions. For instance, the absence of a positive $\lambda$ for $\Nav \sim 0.45$ indicates that the leading solution in this region is a higher order solution. For the more general solution we determine the leading solution to be a higher order $s^*$-wave in this case, as we shall see below.
\begin{figure}[t!]
\centering
 	\includegraphics[angle=0,width=0.45\textwidth]{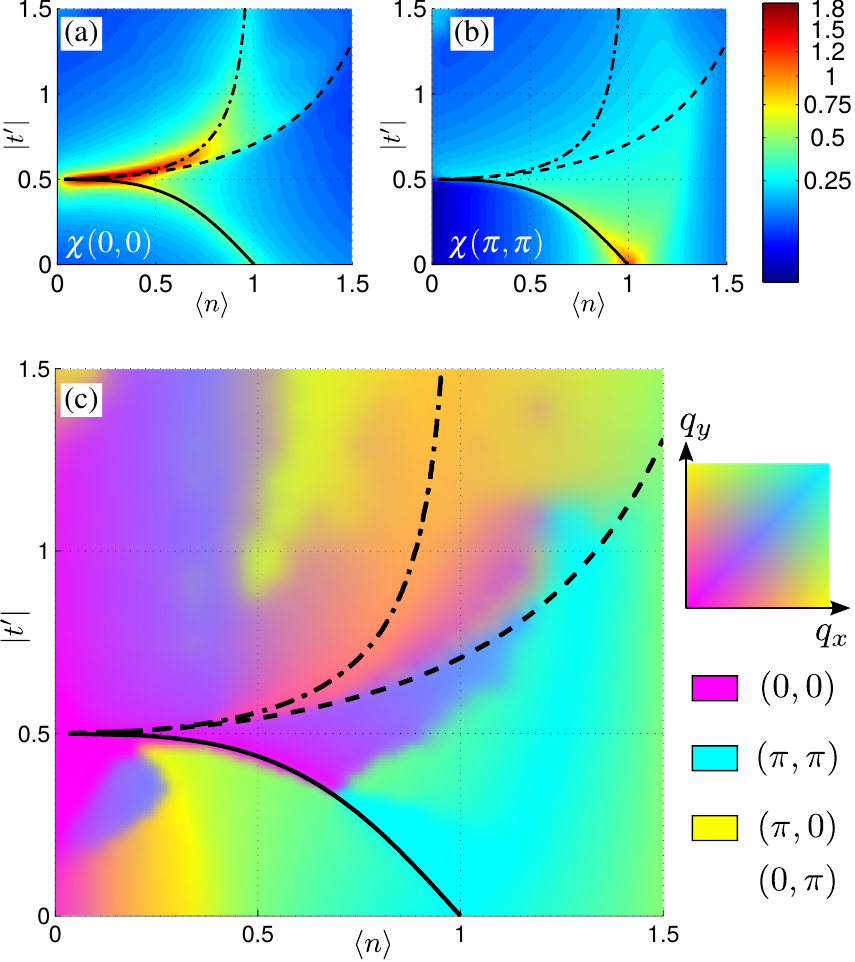}
 	\caption{(Color online) (a) Spin susceptibility Re $\chi_0(\qv,\omega=0)$  at the wave vector $\qv=(0,0)$ as a function of filling, $\Nav$ and next-nearest hopping constant $t'$.
 	The full black line and the black dashed-dotted line show where the van Hove singularity crosses the Fermi level at the positions $(\pm \pi,0)$/$(0,\pm\pi)$ and $(k,\pm k)$, with $k=\pm \cos^{-1}(\frac{1}{2|t'|})$, respectively. The dashed line shows where a hole pocket is removed from the Fermi surface.
 	(b) Spin susceptibility weight at the wave vector $\Qv=(\pi,\pi)$. (c) The wave vector $\qv$, for which the bare spin susceptibility achieve its maximum value (color indicates $\qv$ according to inset),
 	plotted as a function of filling $\Nav$ and next-nearest hopping constant $t'$.}
 	\label{fig:chidiagram}
\end{figure}

Close to half filling the $d_{x^2-y^2}$ solution is clearly dominant. When the system is hole doped away from half filling, the $d_{x^2-y^2}$ solution becomes increasingly strong as the the van Hove critical density, $\Nav_{vH}=0.66$, is approached. However, an abrupt change occurs very close to the van Hove critical density, where the $d_{x^2-y^2}$ solution becomes unstable. This is manifested by a sharp dip of the dashed cyan curve in  Fig.~\ref{fig:lgeProjected0p35}(u).
The $d_{x^2-y^2}$ solution becomes unfavorable due to the development of a peak at $\qv=(0,0)$ in the susceptibility. This gives rise to a large repulsive interaction between neighboring momenta $\kv$ and $\kv'$ in the singlet channel, and causes a suppression of all singlet solutions, in this case the  $d_{x^2-y^2}$ solution. Such a suppression of singlet superconductivity due to the $\qv=(0,0)$ peak in the spin susceptibility was originally discussed by Berk and Schrieffer~\cite{berkschrie} in their pioneering work on spin-fluctuation mediated pairing. As clearly visible in Fig.~\ref{fig:lgeProjected0p35}(u) the sharp dip of the singlet solution around the van Hove critical density is accompanied by an increase in the triplet solution $p'$ shown by the red line. Thus, the $\qv=(0,0)$ susceptibility peak not only suppresses the singlet solution, but actually supports the development of a triplet gap because it gives rise to an effective attraction for neighboring $\kv$ and $\kv'$ at the Fermi surface. In the absence of additional structures in the spin susceptibility the $p$-wave solution is favored, as in the case of very small filling shown in the inset of Fig.~\ref{fig:lgeProjected0p35}(u). However, if the spin susceptibility shows additional peaks as in the case of fillings close to the van Hove critical density, Fig.~\ref{fig:lgeProjected0p35} (d), higher order triplet solutions will be favored.
The detailed structure of the potential turns out to favor a six node $p'$-wave gap, see Eq.~(\ref{eq:f}) and Fig.~\ref{fig:tripletgaps}(b). We will return to this is section~\ref{sec:triplet}.

\begin{figure}[b!]
\centering
 	\includegraphics[angle=0,width=\columnwidth]{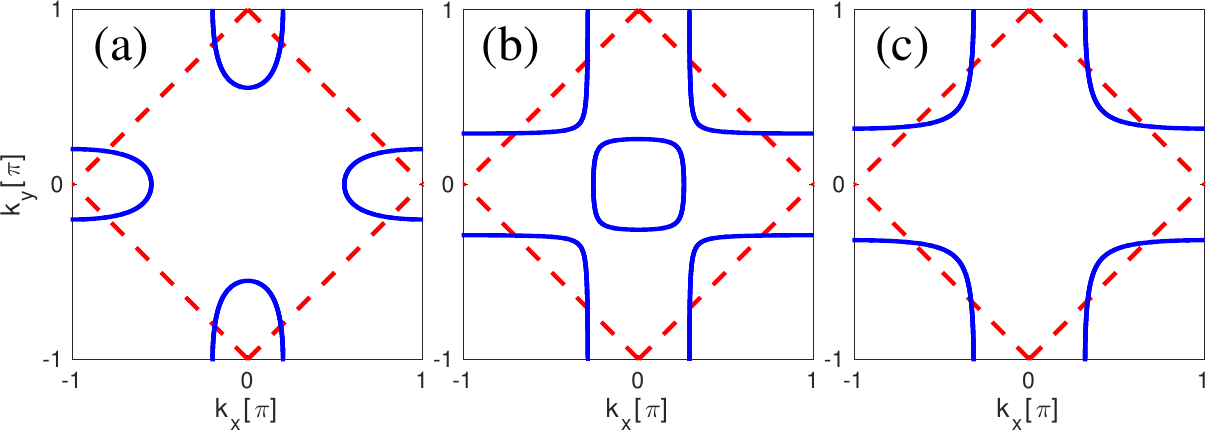}
 	\caption{Fermi surfaces for $t'=-0.8$ at three fillings. (a) At small fillings, $\Nav=0.3$, the Fermi surface consists of four electron pockets centered at $(\pm \pi,0)$ and $(0,\pm \pi)$. (b) For a filling of $\Nav=0.9$, which is above the van Hove critical density, a hole pocket is centered at $(0,0)$ and four hole pockets are centered at $(\pm \pi,\pm\pi)$ and $(\pm \pi,\mp\pi)$. (c) At a larger filling, $\Nav=1.14$, the central hole pocket is removed from the Fermi surface.}
 	\label{fig:FS_largent}
\end{figure}

\begin{figure*}[t!]
\centering
 	\includegraphics[angle=0,width=0.9\linewidth]{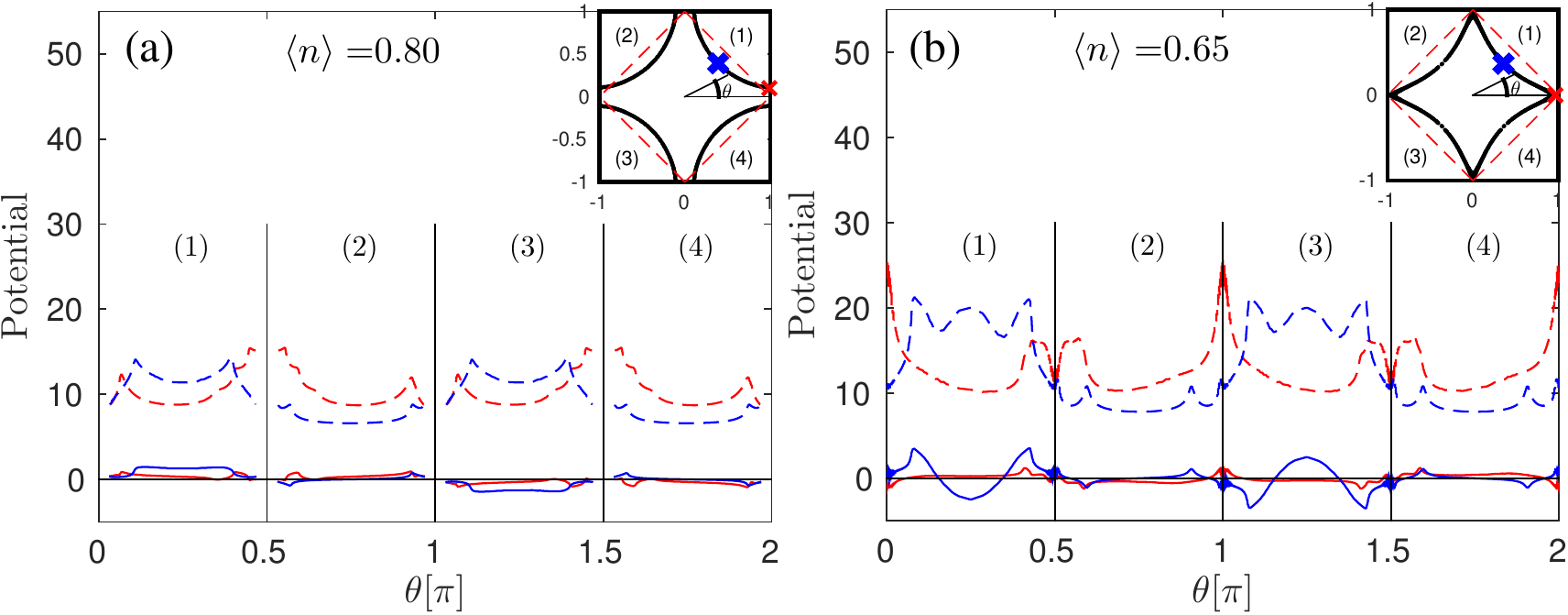}
 	\caption{(Color online) Singlet (dashed lines) and triplet (full lines) pairing interaction in the form stated in Eq.~(\ref{eq:SCGapEquationNS}), $\Gamma^{s/t}_{k,k'} \pm \Gamma^{s/t}_{-k,k'}$ for filling (a) $\Nav =0.80$ and (b) $\Nav=0.65$. Two different positions of the momentum $\kv$ has been chosen, as shown by the colored crosses in the Fermi surface inset. The pairing between $\kv$ and all $\kv'$ in region $(1),(2),(3)$ and $(4)$ of the Fermi surface are shown in the main panels with the position of $\kv'$ being parametrized by the angle $\theta$ measured with respect to the $x$ axis. The next-nearest neighbor hopping is $t'=-0.35$ and the Coulomb interaction is $U=1.75$.}
 	\label{fig:PotVsTheta}
\end{figure*}

\subsection{Phase diagram as a function of doping and $t'$}

In order to further investigate the generality of the results obtained by projection in Fig.~\ref{fig:lgeProjected0p35} we turn to a solution of the
(non-projected) linearized gap equation, Eq.~(\ref{eq:lge}), for different next-nearest neighbor hopping strengths $|t'|$ in the range $0-1.5$.  In contrast to the projection method, this general procedure determines the leading solution amongst all possible higher order harmonics. We classify the solution according to its transformation properties into one of the four singlet solutions $s^*$, $d_{x^2-y^2}$, $d_{xy}$, $g$ or triplet. In the two subsequent sections, we will discuss the triplet states in more detail.  Note that  in general solutions may correspond to  higher order harmonics and have additional nodes compared to  the leading harmonics given in Eqs.~(\ref{eq:s}-\ref{eq:p}).
For each point in the phase diagram $(\Nav, t')$ we solve Eq.~(\ref{eq:lge}) and adjust the value of $U$ such that the leading eigenvalue is $\lambda=0.1$. The result of this procedure is shown in Fig.~\ref{fig:tp-xdiagram}, where the leading superconducting instability for fillings and next-nearest neighbor hopping strengths in the range $0-1.5$ is shown. This procedure is justified as follows.
Since we cover a large range of different Fermi surface structures, a fixed value of the Coulomb interaction of e.g. $U=2$ will cause a break-down of the paramagnetic RPA formalism due to the instability to long range magnetic order. By allowing for a variation in $U$ this is avoided, and at the same time we discuss only
instabilities with a non-negligible critical temperature. Note that this approach is different than a previous report,~\cite{Hlubina99} where the genuinely weak-coupling approach $U \to 0$ limit was taken. Nevertheless, our results show some qualitatively agreement with Ref.~\onlinecite{Hlubina99}; a $g$-wave and a (small) $d_{xy}$ region appear in the filling regime of $\Nav=0.25-0.4$ for $|t'|<0.5$ and at fillings around $\Nav=0.7$ a $d_{x^2-y^2}$ region dominates at all $|t'|<0.4$, whereas an $s$-wave domain takes over for $0.4<|t'|<0.5$ in this filling regime. In our approach, the region of triplet solutions has substantially shrunk compared to Ref.~\onlinecite{Hlubina99}, with different higher order singlet solutions taking over in the regime of small $t'$ close to a filling of $\Nav=0.5$.
In the case of large electron dopings, where the spin susceptibility shows only very weak structure, $U$ must be very large to obtain a leading instability of $\lambda=0.1$ and we omit this regime in Fig.~\ref{fig:tp-xdiagram}. Further, regions for which $U>3$ the colors are less saturated to indicate this.

The structure of the spin susceptibility plays a decisive role for the gap symmetry, and this is very tightly connected to the geometry of the Fermi surface.
Two different regimes of $t'$ give rise to very different Fermi surface geometries.
For $|t'|<0.5$ the Fermi surface evolution as a function of doping is similar to the case $t'=-0.35$ shown in Fig.~\ref{fig:lgeProjected0p35}(k-t). In this case, a transition of the Fermi surface occurs when the van Hove singularities at $(\pm \pi,0)$ and  $(0,\pm \pi)$ cross the Fermi level. This happens when $\mu=4t'$. For $|t'|>0.5$ the transition occurs at $\mu=1/t'$ and invokes van Hove singularities at diagonal positions $k_y=\pm k_x = \pm \cos^{-1}(\frac{1}{2|t'|})$.
The Fermi surface evolution with doping for $t'=-0.8$ is depicted in Fig.~\ref{fig:FS_largent}.
For the special case of $|t'|=0.5$ the van Hove singularity resides at the bottom of the energy band.

The correlation between the van Hove singularity at the Fermi level and the appearance of spin susceptibility weight at the wave vector $\qv=(0,0)$ is shown in Fig.~\ref{fig:chidiagram}(a). Here the value of the spin susceptibility at $\qv=(0,0)$ is shown as a function of $\Nav$ and $t'$. We also show where the van Hove singularity crosses the Fermi level, indicated by the full black line in the case of $|t'|<0.5$) and the dashed-dotted line for the $|t'|>0.5$ case. The black dashed line shows where a hole pocket is removed from the Fermi surface.
In Fig.~\ref{fig:chidiagram}(b) the spin susceptibility weight at the wave vector $\Qv=(\pi,\pi)$ is depicted. As expected, the signatures of a strong peak at $\Qv$ is clearly visible near half filling and $t'=0$. The weight at $\Qv$ expands in  a doping region around half filling for $|t'| < 0.5 $. In addition, it shows a clear correlation with the van Hove critical density shown by the full black line. This explains why the $d_{x^2-y^2}$ solution is increasingly favored upon hole doping away from half filling for $t'=-0.35$, as observed in Fig.~\ref{fig:lgeProjected0p35}(u).

By a comparison of the red regions in Fig.~\ref{fig:tp-xdiagram} and Fig.~\ref{fig:chidiagram}(a), we observe that the two triplet solution branches that expand from the low filling regime in Fig.~\ref{fig:tp-xdiagram} are explained by the weight at $\qv=(0,0)$ in the spin susceptibility, which is correlated with the van Hove critical densities. This is in agreement with the expectations that a $\qv=(0,0)$ peak in the pairing interaction favors triplet superconductivity.
The reason why we see a shift from triplet to singlet ($d_{x^2-y^2}$) superconductivity at the lower red branch around $\Nav \sim 0.6$ is because the $\Qv$ peak becomes dominant in this regime. This transition from triplet to $d_{x^2-y^2}$ superconductivity is most clearly visible in Fig.~\ref{fig:chidiagram}(c), where we capture
the dominating structure of the spin susceptibility at every position $(\Nav, t')$ of the phase diagram, by plotting the wave vector $\qv$ for which the bare spin susceptibility achieve its maximum value. Note how the structure of the spin susceptibility transforms from being dominated by a $\qv \sim (0,0)$ peak in the regime of the triplet branch, to being dominated by the $\Qv$ peak.
Also, from the large magenta region in Fig.~\ref{fig:chidiagram}(c) we see why triplet superconductivity governs the phase diagram in an extended regime around $|t'|=0.5$ for small fillings. As expected, a $\Qv$ peak dominates in the region around half filling.

In other regions of the phase diagram where we observe a different type of singlet superconductivity, other $\qv$ structures of the susceptibility become dominant, as seen in Fig.~\ref{fig:chidiagram}(c). Note, however, that subdominant features in the susceptibility which might influence the gap equation, are not visible from this figure.
The large region of $d_{xy}$ superconductivity, which occurs at small to moderate hole dopings and small electron dopings for $|t'| >0.5$ is correlated with spin susceptibility peaks near $(\pi,0)$ and $(0,\pi)$. We return to the different manifestations of the $d_{xy}$ for small and large $|t'|$ in section~\ref{sec:singlet}.

\subsection{Triplet gap at the van Hove critical density}
\label{sec:triplet}
In the section above we saw how the suppression of singlet superconductivity and concurrently, the development of a triplet gap is intimately related to a $\qv=(0,0)$ peak in the susceptibility which occurs at the van Hove critical density.
Now we turn to a more detailed investigation of the structure of the pairing potential and the consequences for the favored gap symmetries.
In our model two contributions are important for the pairing structure: 1) the hot spot effect, which for most filling levels is more accurately described as a plateau around $\Qv$ rather than a sharp peak at $\Qv$ as seen e.g. from
Fig.~\ref{fig:lgeProjected0p35}(f), 2) the van Hove effect with pairing contributions arising due to the appearance of a quartet of peaks at the diagonal corners at $\qv_{vH}=(\delta,\delta)$ where $\delta \to 0$ as the van Hove singularity crosses the Fermi level (in the case of $|t'|<0.5$). The peaks at $\qv_{vH}$ are visible in Fig.~\ref{fig:lgeProjected0p35}(e) and appear as a purple region in Fig.~\ref{fig:chidiagram}(c) when hole doping is increased towards the van Hove critical doping, shown by the full black line in Fig.~\ref{fig:chidiagram}(c).
 To visualize the hot spot effect and the van Hove effect explicitly, we plot the pairing potential for the band with $t'=-0.35$ and fillings $\langle n\rangle=0.80$ and $\langle n\rangle=0.65$ in Fig.~\ref{fig:PotVsTheta}(a) and \ref{fig:PotVsTheta}(b), respectively. In the case of $\langle n \rangle =0.80$, the largest pairing potential is found due to the hot spot effect, as seen by the red curve close to the angles $\theta=\frac{\pi}{2}$ and $\theta=\frac{3\pi}{2}$ in Fig.~\ref{fig:PotVsTheta}(a). A smaller signature due to the van Hove effect is seen in the red curve close to $\theta=0$ and $\theta=\pi$. As the hole doping is increased, the van Hove effect becomes more pronounced, and very close to the
 van Hove critical density at $\Nav_{vH}=0.66$, we observe sharp peaks at $\theta=0$ and $\theta=\pi$ in the red dashed curve in Fig.~\ref{fig:PotVsTheta}(b). These peaks are responsible for suppression of the singlet solution. At the same time the van Hove effect gives rise to attractive potentials in the triplet channel, as seen in from the full blue line at $\theta=\frac{\pi}{4}$ and the full red line at $\theta=0$.
\begin{figure}[t!]
\centering
 	\includegraphics[angle=0,width=0.9\columnwidth]{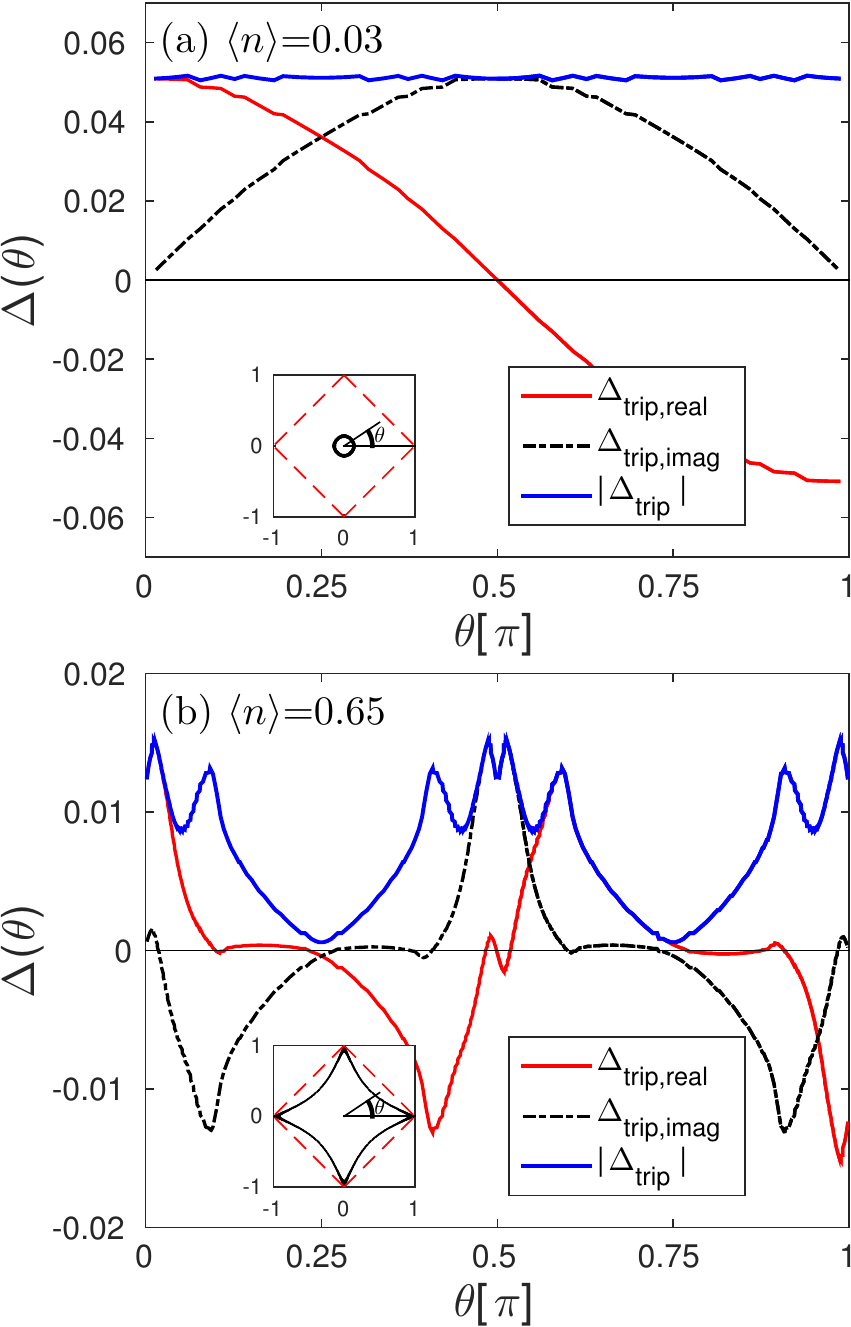}
 	\caption{(Color online) (a) Solution to the full gap equation Eq.~(\ref{eq:SCGapEquationNS}) with triplet potential and the bare Coulomb interaction $U=7.4$ renormalized to $\bar U=3.7$. The filling level is $\Nav=0.03$. Real (full red line) and imaginary (black dashed-dotted line) part of the chiral $p$-wave gap as a function of angle $\theta$ defined in the Fermi surface inset. (b) Solution to Eq.~(\ref{eq:SCGapEquationNS}) with triplet potential and bare Coulomb interaction $U=5.7$ renormalized to $\bar U=1.9$. The filling level is $\Nav=0.65$. Real (full red line) and imaginary (black dashed-dotted line) part of the chiral higher order triplet gap as a function of angle $\theta$ defined in the Fermi surface inset. The absolute value of the gap $|\Delta(\theta)|$ is shown by the blue dash dotted line. In both cases, the next-nearest neighbor hopping is $t'=-0.35$, temperature is $k_BT=0.01$ and energy cut-off is set to $\epsilon_c=0.012$. The size of $\epsilon_c$ does not change the gap structure qualitatively.}
 	\label{fig:TripletGapVsPhi}
\end{figure}

 At very small fillings, the triplet solution has  the simple $p$-wave form  as stated in Eq.~(\ref{eq:p}), but away from the small filling regime, the triplet solution is represented by higher order harmonics. One of these solutions is the $p'$ solution given in  Eq.~(\ref{eq:f}). The main difference between the $p$-wave and the $p'$-wave is that
the latter gap has the same sign for $\textrm{sgn}[\Delta_{\kv+\Qv}]=\textrm{sgn}[\Delta_\kv]$, whereas the $p$-wave solution obeys $\textrm{sgn}[\Delta_{\kv+\Qv}]=-\textrm{sgn}[\Delta_\kv]$. These properties are illustrated in Fig.~\ref{fig:tripletgaps}. When the susceptibility has a peak or a plateau at $\Qv=(\pi,\pi)$, it is favorable for the triplet gap to display the same sign at $\kv$ and $\kv'$ displaced by $\Qv$, since the triplet pairing potential contains the term $-\frac{U^2}{2}\chi(\kv-\kv')$. Therefore the $p'$-wave gap will be favored in the filling regime where the susceptibility shows a $\qv=(0,0)$ peak as well as a peak or plateau structure around $\Qv=(\pi,\pi)$. In the more general case of a quartet peak structure around $\Qv$ as seen in Fig.~\ref{fig:lgeProjected0p35}(d) the triplet gap will resemble the structure of $p'$, but with the nodes slightly displaced, which we will discuss in the next section.

In Ref.~\onlinecite{Guinea04} the changes in gap symmetry as a function of electron- and hole doping within spin fluctuation mediated pairing to second order in $U=6t$ were discussed. The pairing potential was $V(\kv,\kv')=U+U^2\chi(\kv+\kv')$, and the next-nearest hopping constant fixed at $t'/t=-0.276$.
The $d_{x^2-y^2}$ solution dominates at all moderate doping levels, also at the van Hove critical density, in agreement with our findings. However, in contrast to our results,  Ref.~\onlinecite{Guinea04} reports a regime of triplet superconductivity at smaller fillings. As mentioned above, Ref.~\onlinecite{Hlubina99} also reports an extended region of triplet superconductivity which is unrelated to the van Hove critical density. We suspect this discrepancy to arise from the details of procedure; in Refs.~\onlinecite{Hlubina99} and ~\onlinecite{Guinea04} gap solutions was truncated to the first 15 harmonics, whereas in our case we do not invoke any restrictions on the gap functions.

In a recent work by Deng {\it et al.},~\cite{Deng14} the emergence of pairing for the paramagnetic liquid was also addressed in a numerical study of the two-dimensional Hubbard model with $t'=0$. They reported a transition from $p$-wave superconductivity at small fillings through a $d_{xy}$ gap at intermediate filling levels to a $d_{x^2-y^2}$ symmetry close to half filling. For small values of $U$, a higher order triplet gap with six nodes was also found for fillings at $\Nav\simeq 0.55$. This triplet solution is thus unrelated to the van Hove critical density, and we do not find a similar solution at $t'=0$ in our case. Their findings, however, agrees with the results in Ref.~\onlinecite{Hlubina99} which report the $U\to 0$ case. The disagreement might arise from the difference in strength of $U$.

\subsection{Time reversal broken triplet gap solutions}
The fact that all triplet solutions found in the linearized approach are two-fold degenerate suggests that a TRSB solution might be favored. Therefore we turn to the full gap equation as given in Eq.~(\ref{eq:SCGapEquationNS}) in the triplet channel, which we solve at $k_BT=0.01$.
We consider two special fillings of $\Nav=0.03$ and $\Nav=0.65$ where in both cases $t'=-0.35$. At very low filling, the susceptibility exhibits only a weak structure around $\qv=(0,0)$ as seen in Fig.~\ref{fig:lgeProjected0p35}(a). In this case the $p$-wave triplet solution is favored and, as shown in Fig.~\ref{fig:TripletGapVsPhi}(a), the full solution is the nodeless TRSB gap of the form $p_x \pm i p_y$. In this filling regime, strong Coulomb interactions are required to achieve a superconducting instability due to the weak structure of the spin susceptibility. We use a bare Coulomb interaction of $U=7.4$ which is renormalized in the RPA expressions to $\bar U=3.7$.

At higher fillings, the spin susceptibility acquires more structure and supports a superconducting gap with a bare Coulomb interaction $U=5.7$ renormalized to $\bar U=1.9$. The preferred solution in this case is also a TRSB solution.
If the TRSB solution had been of the form $p'_x \pm i p'_y$ only the nodes along the zone axes $k_x$ or $k_y$ would be lifted with the four nodes along the zone diagonals preserved. In this case, since only two of the six nodes are lifted, the gain in condensation energy of the TRSB solution compared to one of the solutions, $p'_x$  or $p'_y$, would be limited.
However, in the present case of $\Nav=0.65$ where the spin susceptibility displays a quartet of peaks around $\Qv$, the simple form $p'_x\pm i p'_y$ is replaced by a more complicated gap solution displayed in Fig.~\ref{fig:TripletGapVsPhi}(b), which indeed provides a fully gapped TRSB solution. From the angular gap dependence shown in Fig.~\ref{fig:TripletGapVsPhi}(b), strong effects due to the susceptibility structure are clearly visible from the absolute value of the gap, $|\Delta_{\rm trip}(\theta)|$ as shown by the full blue line. The maximum of the gap achieved close to $\theta=0$ and $\frac{\pi}{2}$ is due to the van Hove effect, and the peaked feature at $\theta \sim \pi/8$ is related to the quartet of peaks around the $\Qv$ vector in the susceptibility as shown in Fig.~\ref{fig:lgeProjected0p35}(d).
This underlines again the strong connection between the Fermi surface structure, spin susceptibility, and the detailed angular dependence of the superconducting gap.
Lastly, we note that the gap minimum of this higher order triplet solution is achieved at $\theta=\frac{\pi}{4}$, which is at the position of the nodal lines of the $d_{x^2-y^2}$ structure. 

\begin{figure}[t!]
\centering
 	\includegraphics[angle=0,width=0.49\textwidth]{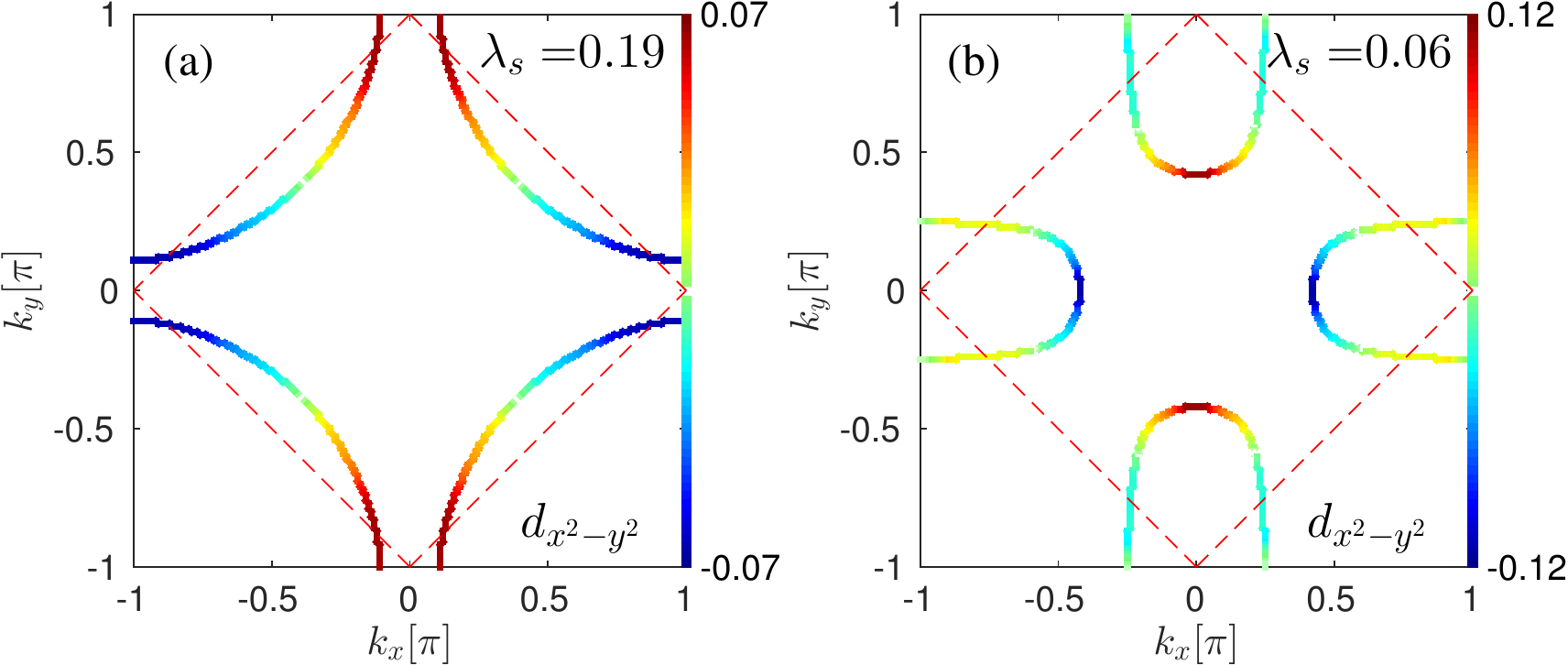}
 	\caption{(Color online) (a) $d_{x^2-y^2}$ solution to the linearized gap equation as stated in Eq.~(\ref{eq:lge}) for a band with $t'=-0.35$ and $U=1.75$ at a filling of $\Nav=0.8$.
 	(b) $d_{x^2-y^2}$ solution to the linearized gap equation as stated in Eq.~(\ref{eq:lge}) for a band with $t'=-0.8$ and $U=1.75$ at a filling of $\Nav=0.5$. The magnetic zone boundary is indicated by the red dashed line.}
 	\label{fig:lged2}
\end{figure}

\subsection{The $d_{x^2-y^2}$ solution}
The $d_{x^2-y^2}$ solution of the one-band Hubbard model is commonly discussed in the region with $|t'|<0.5$ around half filling, due to the relevance for cuprates.
From the phase diagram in Fig.~\ref{fig:tp-xdiagram}, we observe that there is in fact another large region of the phase diagram for which a $d_{x^2-y^2}$ solution dominates corresponding to  $|t'|>0.5$ where the Fermi surface topology is quite different. In order to show a solution in both regimes, we plot in Fig.~\ref{fig:lged2} the $d_{x^2-y^2}$ solution in the case of $t'=-0.35$, $\Nav=0.8$ and $t'=-0.8$, $\Nav=0.5$. Note that even though the Fermi surface of the $t'=-0.8$ band has no Fermi surface weight along the zone diagonals, which are the nodal lines of the simple harmonic $d$-wave, it displays more nodes at the Fermi surface than the solution for the $t'=-0.35$ band, namely eight nodes instead of four.
 \begin{figure}[t!]
 \centering
  	\includegraphics[angle=0,width=0.99\linewidth]{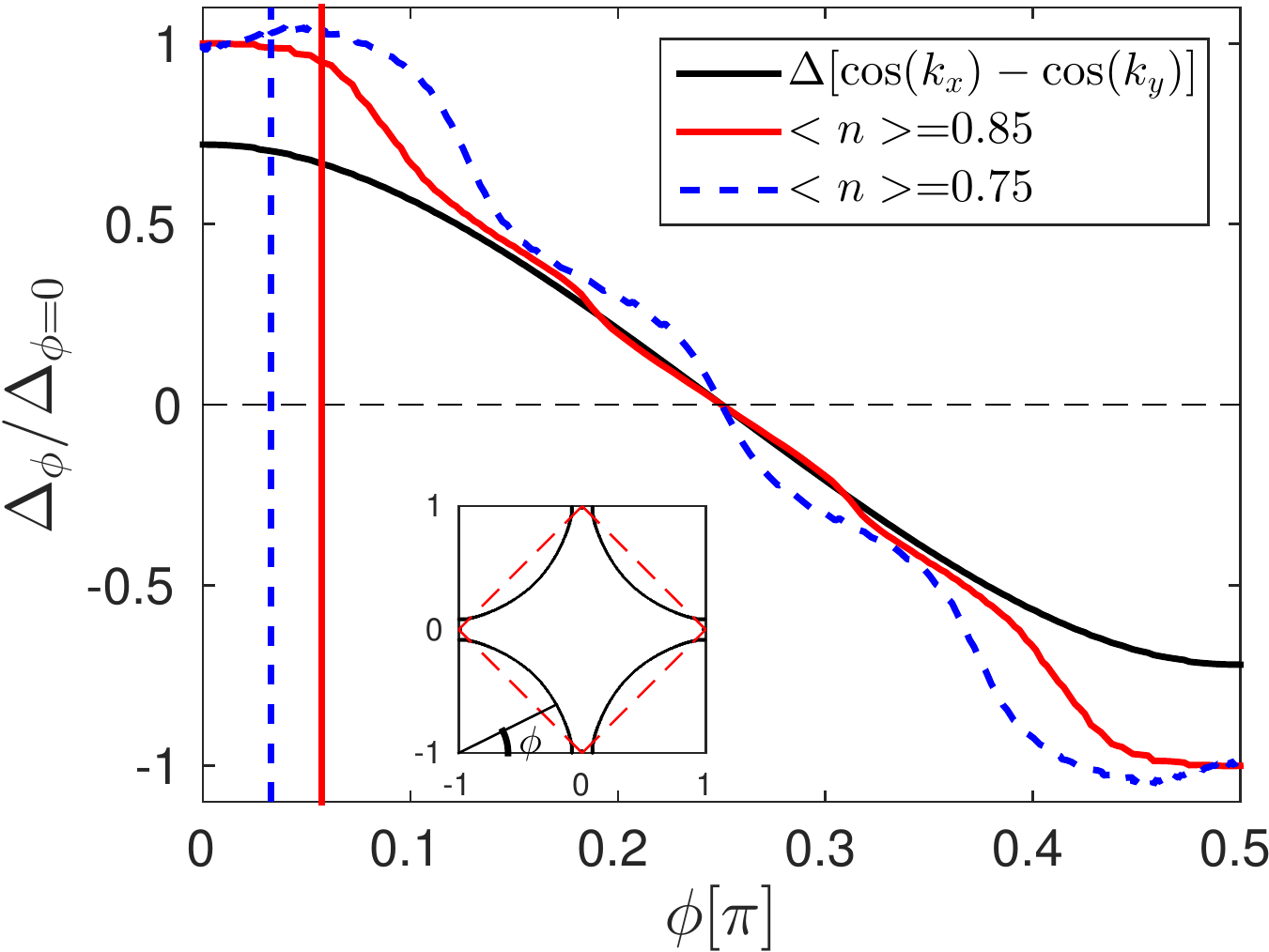}
  	\caption{(Color online) Solution to the full gap equation Eq.~(\ref{eq:SCGapEquationNS}) with singlet pairing interaction for $t'=-0.35$ at filling levels
  	$\Nav=0.85$ (full red line) and $\Nav=0.75$ (dashed blue line) at temperature $k_BT=0.01$. The renormalized Coulomb interaction is $\bar U=2,1.88$, respectively and the Coulomb renormalization is $z=3$. The energy cut-off is set to $\epsilon_c=0.015$, and the gap value at the antinodes is $\Delta_{\phi=0}=0.01$ in both cases.
  	The angle $\phi$ is defined in the Fermi surface inset. The vertical lines show the position of the hot spots for the $\Nav=0.85$ (full red line) and $\Nav=0.75$ (dashed blue line). Note that there is a clear correlation between the position of the hot spot and enhancement of the superconducting gap away from a simple harmonic gap function in the case of $\Nav=0.85$. However,
  	at $\Nav=0.75$, which is closer to the van Hove critical density, the higher harmonic angular dependence of the gap function is not directly related to the position of the hot spot.}
  	\label{fig:SingletGapVsTheta}
 \end{figure}

A closer inspection of the $d_{x^2-y^2}$ solution for the band with $t'=-0.35$ indicates deviations from the simple form of $\Delta [\cos k_x -\cos k_y]$ due to the presence of higher harmonics, i.e. longer range superconducting pairing interaction. The maximum gap value is not achieved at the antinodal points, but shifted towards the nodal direction. This effect has been discussed previously as a signature of spin-fluctuation-mediated pairing.~\cite{Blumberg02,Khodel04,Guinea04,Parker08,Eremin08,Krotkov06} In previous work special attention was drawn to the hot spot effect in which the gap maximum occurs at the $\kv$ position of the hot spot.
In Fig.~\ref{fig:SingletGapVsTheta} we show the singlet gap as calculated by the full (non-linearized) gap equation, Eq.~(\ref{eq:SCGapEquationNS}).
In the figure we also show the position of the hot spot, i.e. the angle at which the Fermi surface intersects the magnetic zone boundary. This is shown by the vertical lines in Fig.~\ref{fig:SingletGapVsTheta}. It is seen that the non-monotonicity of the gap is not directly related to the hot spot effect, since the hot spot position  moves towards $\phi=0$, but the strong gap enhancement moves closer to the nodal direction
upon increased hole doping. Close to the van Hove critical density, the higher harmonic content of the gap function becomes more pronounced. This tendency was also pointed out in Ref.~\onlinecite{Guinea04}.
Note that proximity of the van Hove singularity leads to an increase in the number of states participating in the formation of Cooper pairs, but the corresponding formation of additional nesting peaks in the susceptibility at small $\qv_{vH}=(\delta,\delta)$, see Fig.~\ref{fig:lgeProjected0p35}(e), in fact work against the $d_{x^2-y^2}$ solution, since in the singlet channel this will favor nodes along the zone axes.

\begin{figure}[t!]
\centering
 	\includegraphics[angle=0,width=0.23\textwidth]{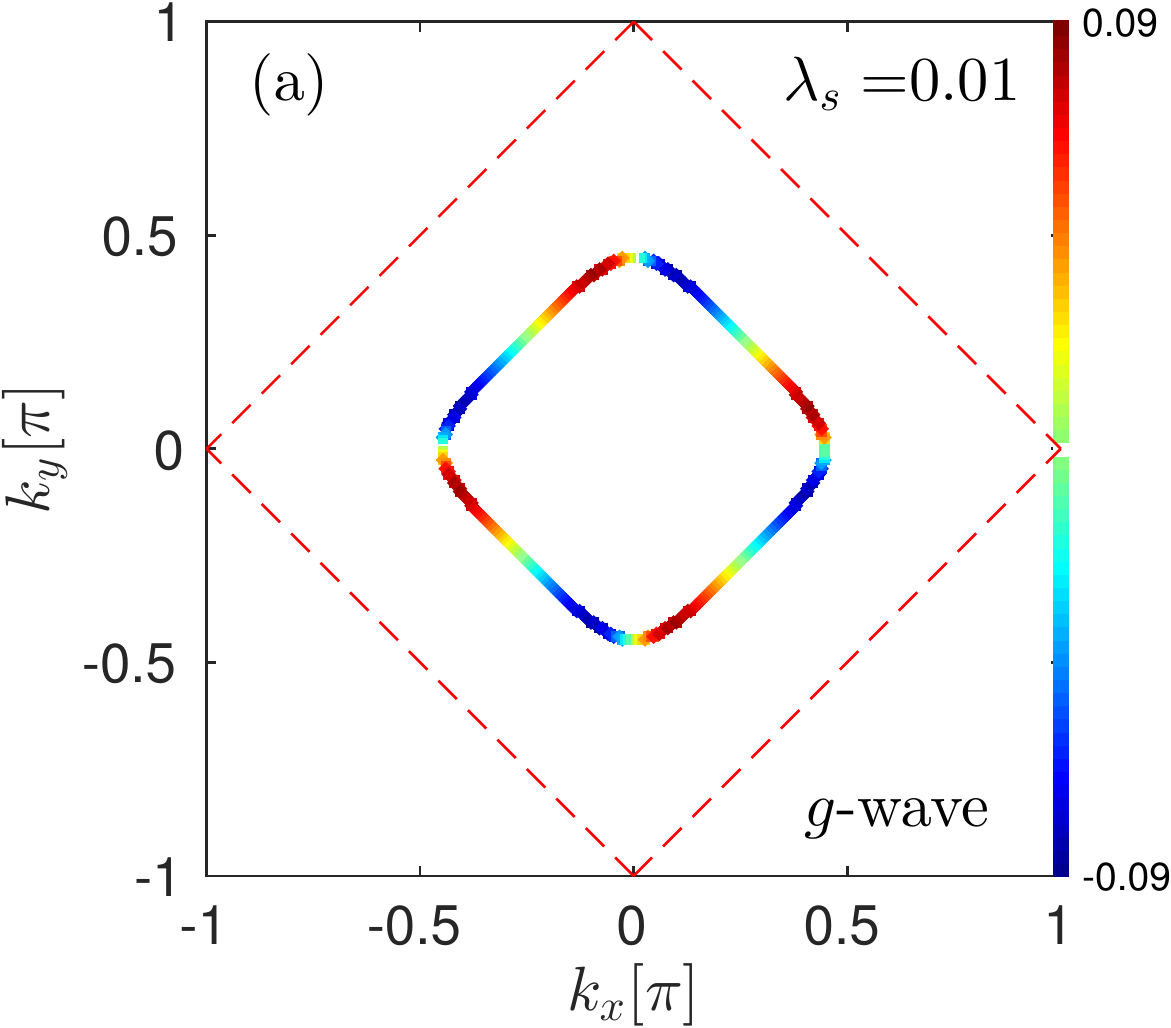}
 	\includegraphics[angle=0,width=0.23\textwidth]{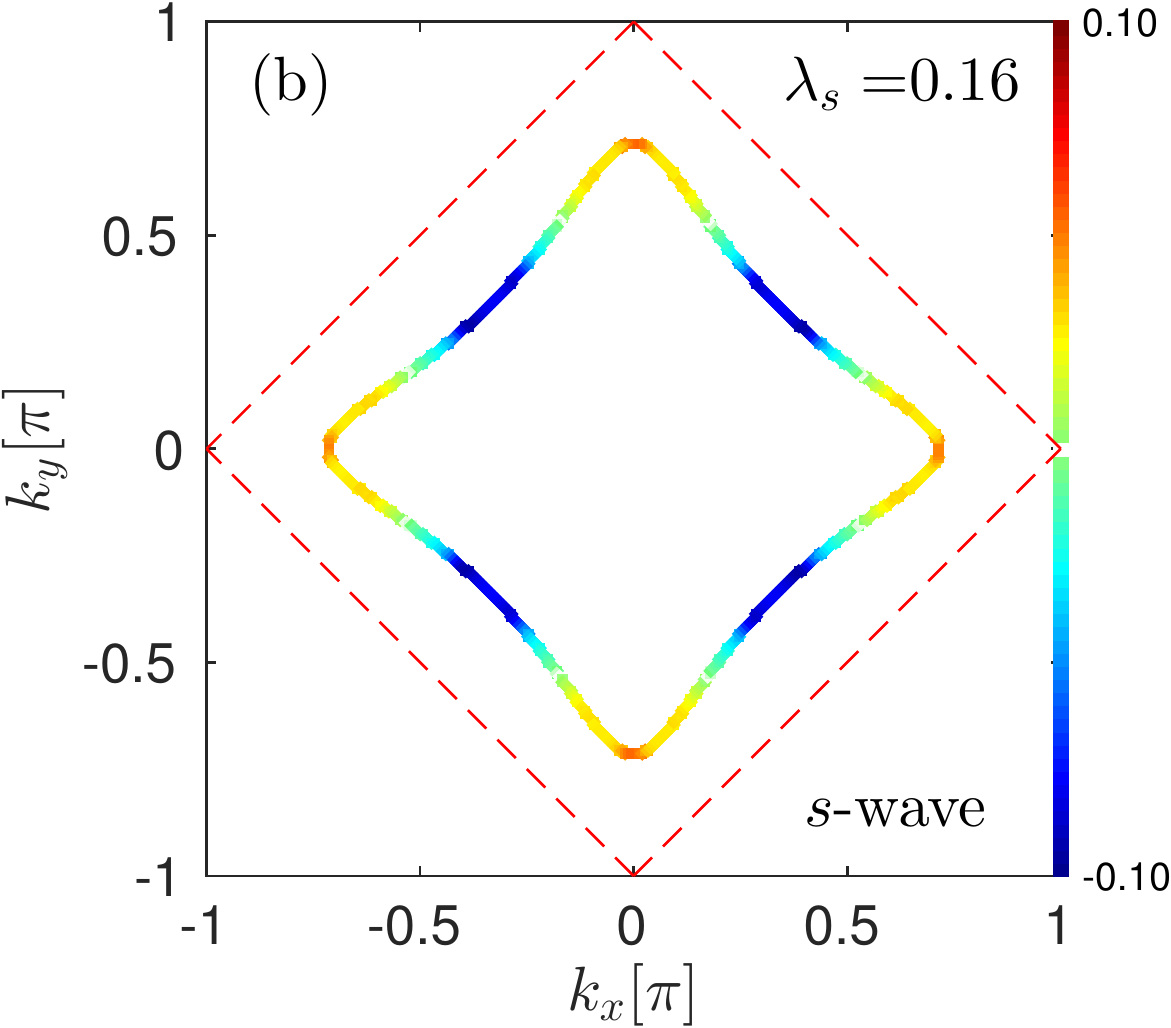}
 	\caption{(Color online) Leading solutions to the linearized gap equation as stated in Eq.~(\ref{eq:lge}) for a band with $t'=-0.35$ and $U=1.75$.
 	(a) At filling of $\Nav=0.25$ the singlet $g$-wave solution dominates.
 	(b) At intermediate filling level of $\Nav=0.50$ the singlet higher order $s$-wave solution with eight nodes is the leading instability. The magnetic zone boundary is indicated by the red dashed line.}
 	\label{fig:lgesing}
\end{figure}

\subsection{Other singlet solutions}
\label{sec:singlet}
\begin{figure}[t!]
\centering
 	\includegraphics[angle=0,width=\columnwidth]{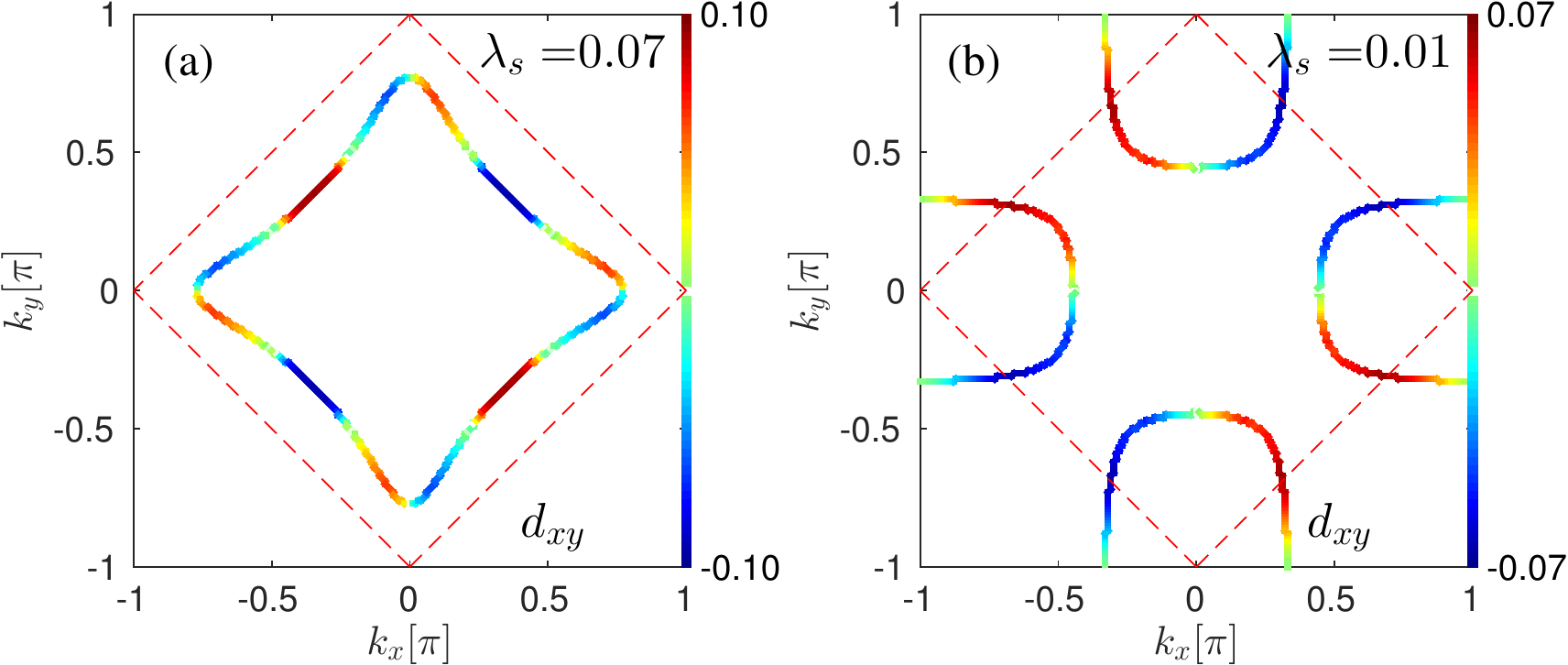}
 	\caption{(Color online) (a) $d_{xy}$ solution to the linearized gap equation as stated in Eq.~(\ref{eq:lge}) for a band with $t'=-0.35$ and $U=1.5$ at a filling of $\Nav=0.55$.
 	(b) $d_{xy}$ solution to the linearized gap equation as stated in Eq.~(\ref{eq:lge}) for a band with $t'=-1.2$ and $U=1.5$ at a filling of $\Nav=0.65$. The magnetic zone boundary is indicated by the red dashed line.}
 	\label{fig:lgedxy}
\end{figure}

From the phase diagram in Fig.~\ref{fig:tp-xdiagram} we observe that the regime of $|t'|<0.5$ and intermediate hole doping levels, have two robust regions of singlet superconductivity besides the $d_{x^2-y^2}$ solution, namely a $g$-wave and $s$-wave region. We show the solutions
for $t'=-0.35$ at fillings $\Nav=0.25$ and $\Nav=0.5$ in Fig.~\ref{fig:lgesing} for which we obtain a $g$-wave, and higher order $s$-wave, respectively.
In the regime of $|t'|>0.5$ the $d_{xy}$ solution is largely dominating in a large region close to half filling. In Fig.~\ref{fig:lgedxy} we show the $d_{xy}$ solution in the case of $|t'|<0.5$ (Fig.~\ref{fig:lgedxy}(a))  and $|t'|>0.5$ (Fig.~\ref{fig:lgedxy}(b)). Due to the difference in Fermi surface topology, the appearance of the $d_{xy}$ solution is quite different in the two cases.

\section{CONCLUSIONS}

In this paper we have studied the superconducting gap structures in a single band Hubbard model within spin-fluctuation-mediated Cooper-pairing scenario in the weak-coupling paramagnetic limit for an extended region of phase space. It complements our earlier study of pairing in the spin density wave phase in the same model.~\cite{WenyaNJP,SDWarxiv} In contrast to previous studies of the paramagnetic phase, our main emphasis was to study the gap structure for a large range of next-nearest neighbor hopping integrals, $t'$, and doping levels away from half-filling, which could be potentially relevant for future systems including new classes of unconventional superconductors as well as optical lattices loaded with interacting fermions. 
We discussed the details of the gap structure and related this directly to the spin susceptibility at all filling levels. Furthermore, we also focused on the role of a van Hove singularity in close proximity to the Fermi level for the transition between various Cooper-pairing channels. This has drastic effects on the gap symmetry since it strongly suppresses singlet superconductivity and leads to the emergence of a nodeless TRSB triplet gap solution. This is a direct consequence of the additional $\qv=(0,0)$ peak structure in the spin susceptibility which reflects the presence of a van Hove singularity at or very near to the Fermi level.

\section*{Acknowledgements}

We thank A.V. Chubukov, and S. Mukherjee for useful discussions. B.M.A., A.T.R. and A.K. acknowledge support from a Lundbeckfond fellowship (Grant A9318).  P.J.H. was supported by NSF-DMR-1005625. I.E. acknowledges financial support of the SPP 1458
“Eisen-Pniktide” of the Deutsche Forschungsgemeinschaft, the German Academic Exchange Service (DAAD PPP USA No. 57051534). T.A.M. acknowledges the support of the Center for Nanophase Materials Sciences, which is a DOE Office of Science User Facility.
The work of M.A.M. was financially supported by a Kazan (Volga region) Federal University grant targeted at strengthening the university's competitiveness in the global research and educational environment. M.A.M. is thankful to the
Ruhr-Universit\"at Bochum for hospitality during the work on
the manuscript.

\end{document}